\documentclass{aipproc}
\layoutstyle{8s}
\newcommand{\dgr}{^\circ}
\newcommand{\lta}{\mbox{\small\raisebox{-0.6ex}{$\,\stackrel
{\raisebox{-.2ex}{$\textstyle <$}}{\sim}\,$}}}
   
\title[{\em Planck} LFI Polarization Systematics]{Systematics of Microwave 
Polarimetry with the {\em Planck} LFI}
\author{J. P. Leahy}{
address = {University of Manchester, Jodrell Bank Observatory, Macclesfield,
Cheshire, SK11 9DL, England},
email = {jpl@jb.man.ac.uk}}
\author{V. Yurchenko}{
address = {Dept. of Experimental Physics, National University of Ireland, 
Maynooth, Co. Kildare, Ireland},
email = {v.yurchenko@may.ie}}
\author{Morag A. Hastie}{
address = {University of Manchester, Jodrell Bank Observatory, Macclesfield,
Cheshire, SK11 9DL, England}}
\author{M. Bersanelli}{
address = {Universita di Milano, Dipartimento di Fisica, via Celoria 16, 
Milano, I-20133, Italy},
email = {marco@ifctr.mi.cnr.it}}
\author{N. Mandolesi}{
address =  {TESRE-CNR, via P. Gobetti 101, Bologna, I-40129, Italy},
email = {mandolesi@tesre.bo.cnr.it}}

\begin{abstract}The {\em Planck} 
Low Frequency Instrument will recover polarization by
differencing the outputs from radiometers sensitive to orthogonal
polarizations. We contrast the systematic errors that afflict
such a system with those that affect correlation polarimeters; the
{\em Planck} design has some important advantages when measuring very weak
signals such as the CMB.  We also review systematic effects
arising from the choice of scan strategy for all-sky mapping
missions like {\em Planck}. [For the Planck-LFI consortium].
\end{abstract}
\begin{document}
\maketitle
\section{Differencing vs. Correlation Polarimeters}
Radiometers can be used to measure polarization in two fundamentally 
different ways.  

Differencing polarimeters
work in much the same way as polarimeters used in other wavebands; that is,
the radiation is filtered to isolate components in several different 
pure polarization states and these are combined to derive the required
Stokes parameters. The {\em Planck} Low Frequency Instrument (LFI), described
in these proceedings by Villa, is of this type.  Each
feed horn couples the incoming radiation to a waveguide, essentially
preserving the polarization state (deviations from
this approximation will be discussed later). 
An OMT separates the radiation into two orthogonal linear components 
(`$X$' and `$Y$') which are separately amplified and square-law detected.
Ideally, the sum of these signals is proportional to Stokes $I$, while
the difference measures one component of the $(Q,U)$ vector.
This is formally equivalent to a Wollaston prism system in optical
polarimetry. To obtain the full linear polarization vector, more information
is needed, and usually a second feed oriented at $45\dgr$ to the first is
used, which also provides a redundant measurement
of $I$. This arrangement leaves $V$ undetermined, but, as we believe there
is no significant background of circular polarization, this is not a drawback
for {\em Planck}.  

Square-law detection systems have offsets due to receiver 
noise, noise from lossy components in the signal path, and of course from
their sensitivity to the dominant unpolarized emission.
The offsets are notoriously variable due to physical temperature changes
and also to amplifier gain changes, and
they will directly affect the measured $Q$, $U$.  For this reason most
modern polarimeters are based on correlation rather than differencing.
Correlation polarimeters make use of phase coherence
to derive the full polarization state of the incoming
radiation from a single horn. Again an OMT is used to separate two
orthogonal polarizations, and each channel is square-law detected. But
in addition, the signals from the two channels are correlated,
i.e. multiplied together,
with and without a $90\dgr$ phase shift. The two correlation products
give the remaining two Stokes parameters.  Because all
parameters are measured with the same feed, this gives an improvement 
of $\sqrt{2}$ in sensitivity to linear polarization over a simple differencing
system with the same number of feeds; and compared to a more practical
null-balanced system (as in the LFI) the correlation receiver has a full
factor of 2 advantage. When
the main interest is in linear polarization, a polarizer (quarter wave
plate equivalent) 
is usually inserted ahead of the OMT to convert circular to linear 
polarization, so that the OMT separates polarizations corresponding
to right- and left-hand circular on the sky. The correlation
products are then the two components of the $(Q,U)$ vector.  

Ideally,
there would be no offsets in the correlation products, 
because zero polarization
implies zero mean correlation between RHC and LHC.  
In reality cross-polarization (failure of the OMT to completely separate
the RHC and LHC signals) and cross-talk (leakage of noise generated in
each amplifier chain into the other) mean that the two channels do have
a correlated component, but in conventional systems this polarization
offset is much smaller than that on the outputs of the square-law
detectors.  A numerical example is instructive.  On a general-purpose 
telescope, the OMT is required to cover a bandwidth of up to 20 or 30\%,
which limits the polarization purity
to typically 25--35 dB; i.e. the cross-polar voltage is reduced by a factor
of  $\sim 30$.  As the corresponding voltage in the other channel is 
unattenuated by definition, the contribution to the correlator offset
is $\sim T_{\rm Antenna}/30$.
In total-power systems the offset is dominated by
$T_{\rm receiver}$ rather than $T_{\rm Antenna}$, but 
in a correlation system we can reduce the cross-talk by more than the 
cross-polarization, by using circulators or equivalent to provide 
an additional $\sim 20$ dB isolation between the channels.
Thus we expect an offset of order $T_{\rm Antenna}/30
+ T_{\rm receiver} /300$. 
Compared to a simple total power system, with an offset of 
$T_{\rm Antenna}+ T_{\rm receiver}$,
this is a very big improvement.  But consider
the goal of measuring $\mu$K polarizations in the presence of typically 
$20$ K of receiver noise, and hence a polarization offset of
$\sim 100$ mK.  Fluctuations of one part in $10^{4}$ would swamp the
wanted signal.

To summarise, in a correlation system, both cross polarization and 
cross-talk give a false positive signal.  Dedicated instruments must therefore
be designed to reduce and stabilise these effects. The former is difficult
because the critical factor is the voltage, rather than the power, of the 
cross term. For instance SPOrt, despite an OMT with 60 dB
isolation (Peverini et al., these proceedings), 
still requires the stability offered by a space mission.

In contrast, we will see that cross-polarization and cross-talk are
rather unimportant in differencing systems.  False positives are instead
generated by the total power offsets.  But these offsets are also,
of course, a serious problem for total power measurements, and practical
systems are designed to minimise their effects.  When extreme care
is taken to do this, as in the {\em Planck} LFI, the dominant polarization
systematic is eliminated at the same time.  One is then in a position to
make accurate measurements of polarization even when the telescope suffers
from cross terms that would be difficult to cope with if it was 
operated as a correlation polarimeter.

\section{The LFI radiometers}

As in any space experiment, the LFI is designed to minimise mass, power
consumption and complexity consistent with its primary science goal,
which is to measure temperature fluctuations in the CMB.  For this
reason there is no correlator, and the
principle reason for measuring both polarizations in each horn is
to gain a factor of $\sqrt{2}$ in sensitivity to $I$. Nevertheless,
the potential to measure polarization has always been recognised, and
the design has been optimised for this where there is no conflict with
the primary goals.

Each feed horn in the LFI feeds two radiometers, one for each
polarization. Each radiometer continuously compares the sky brightness 
with that of an internal 4 K reference load.  As described by Villa
(these proceedings) the design ensures that the gain (including
$1/f$ fluctuations) is identical for the sky and load 
signals up until detection.
After detection, the load gain is adjusted by $\sim$10\% 
to account for the known difference between the load and mean sky
system temperatures, and the two signals are differenced to yield a 
(nominally) zero mean output, 
whose amplitude is generally dominated by the CMB dipole and 
therefore is typically $I - T_0 \sim$2 mK.   Despite all these precautions the
output is subject to $1/f$ noise mainly due to variations in the 
amplifier noise temperature, but also due to 
fluctuations in the reference load temperature, etc. 
However, the amplitude of of the $1/f$ noise is designed to be 
below that of the thermal noise for frequencies greater than the 1 rpm 
spin period of {\em Planck}.  Each feed scans the same
circle on the sky, 60 times over between hourly repointings, and
stacking the data effectively filters out noise except at frequencies very
close to harmonics of the spin frequency. Thus the low frequency $1/f$ 
noise contributes only to the zero harmonic, a single undetermined
offset on each scan circle.  Both offset and the average gain can be
calibrated by matching to an assumed CMB dipole, after excluding 
(or modelling) Galactic plane emission \citep{Bersanelli1997}. 
Note that errors in
the assumed dipole will cause identical gain errors in the two polarization
channels, which will not affect the polarization signal to first order.
In practice, residual offsets and gain errors are expected, due to 
errors in the assumed dipole direction, imperfect Galactic masking, 
and residual $1/f$ noise at low order harmonics, but the multiply-redundant 
coverage of the sky means that these can be removed by de-striping techniques
\citep{Maino1999},
which can even be applied directly to the polarization signal 
\citep{Revenu2000}.

The advantages of the LFI design for polarization measurements
could be considerably enhanced if we jettisoned the aim of measuring
the total intensity.  
In this case we could compare the $X$ and $Y$ polarizations
with each other directly instead of with the 4 K loads, so that the
output of the radiometer was a direct measure of a linear polarization
component.  This has several advantages. Most importantly the system
temperatures of the two channels would be matched at a level of $\sim 0.1$ K,
instead of $\sim 1$ K, giving better nulling of the residual $1/f$ noise.
In addition, the thermal noise is reduced
by $\sqrt{2}$ because we have eliminated one level of differencing. Finally,
the complexity of the instrument is sharply reduced by eliminating the
reference loads, half of the amplifier chains, and half the detectors.
Such a design may be appropriate for the post-{\em Planck} era, when
the focus on CMB research will be on polarization.

\subsection{Polarization response of the LFI radiometers}
From the above discussion, we can write the power detected by a given
radiometer, say `$P$', as
\begin{equation} 
s_P = g_P [ T_P - T_0 + R_P + N_P ]
\label{power}
\end{equation}
where $g_P$ is the gain, $T_P$ is the antenna temperature
\citep{Berkhuijsen1975} in the polarization 
matched to this radiometer, $T_0$ is the 
temperature assumed for balancing, $N_P$ is white noise and $R_P$ is
red (or $1/f$) noise.  After averaging the data into rings, 
$N_P$ is independent at each pixel but $R_P$ is highly correlated. 

The polarized brightness can be written in several ways. One of the most
useful is in terms of sensitivities to the four Stokes parameters:

\begin{equation}
T_P = {1 \over 4 \pi}
\int ({\cal I}_P I + {\cal Q}_P Q + {\cal U}_P U + {\cal V}_P V) \, d\Omega 
\end{equation}
where ${\cal S}_P$ is the beam of radiometer $P$ in Stokes parameter $S$
(measured in terms of brightness temperature).
Taking $V = 0$, we can re-write this as
\begin{equation}
T_P = {1 \over 4 \pi}
\int {\cal I}_P \{I + \epsilon_P(Q \cos 2\phi_P + U \sin 2\phi_P) \} \, d\Omega 
\label{TP}
\end{equation}
where $\epsilon_P$ is the efficiency of linear polarization response 
and $\phi_P$ is the orientation of peak response.
Inserting Eq.~\ref{TP} into Eq.~\ref{power}, 
we note the presence of
a large total intensity term $I - T_0$, of order $10^3$ times the polarized
term, as previously discussed.

The two radiometers in each feed are nearly orthogonal so
we write $\phi_Y \approx \phi_X + \pi/2$. Thus if we difference
the calibrated outputs we get a polarization signal

\begin{equation}
P = {1 \over 2} \left({s_X \over g'_X} - {s_Y \over g'_Y} \right)\approx 
 \epsilon(Q \cos 2\phi + U \sin 2\phi)
\end{equation}
where $g'$ is our estimate of the true gain $g$ 
(for simplicity we idealise the beam as a point measurement for now). 

More precisely, let
$g/g' = 1 + \delta g$, let there be an error $\delta\phi$ in the
orthogonality of the two radiometers, and let 
$\epsilon_X -\epsilon_Y = \delta \epsilon$. Then to second order in the
error terms, 
\begin{eqnarray}
P &=& \langle\epsilon\rangle (Q \cos 2\phi + U \sin 2\phi) \, 
\left[1 + \langle \delta g \rangle - {\delta\phi^2 \over 2}
+ {(\delta g_X - \delta g_Y) \delta\epsilon \over 4 \langle\epsilon\rangle } 
\right] + {\delta g_X - \delta g_Y \over 2}(I - T_0) \nonumber \\
& & + \: \langle\epsilon\rangle
(Q \sin 2\phi - U \cos 2\phi) \, \delta\phi \,
\left[ {\delta g_X - \delta g_Y \over 2} + 
       {\delta\epsilon \over 2\langle\epsilon\rangle} \right]
+ {R_X - R_Y + N_X - N_Y \over 2}
\label{error_terms}
\end{eqnarray}
Unlike the situation for a correlation polarimeter, cross-polarization 
(represented here by $\delta\phi$ and $1 - \epsilon$) 
only appears in the second order, (except for the overall scaling by 
$\langle\epsilon\rangle$), 
and even then, only affect terms proportional to $Q$ and $U$, not $I$.
Thus, even if these error terms were completely unknown, we could still
detect polarization. In reality we expect them to be very small and
calibratable, allowing accurate measurement of the polarized background.

\section{LFI Polarization systematics}
\subsection{Overview}
The {\em Planck} consortium has chosen to analyse systematics in terms
of a number of sources, all of which impact on both total intensity
and polarization measurements.
Many are common in origin and/or effect between HFI and LFI and are
being studied jointly.  These headings and their principle polarization
impacts on the LFI are as follows:
\begin{description}
\item[Far Sidelobes:] Contamination of the polarized signal by total
intensity straylight from the Sun and  Galaxy
(because the ${\cal I}_P$ far sidelobes are 
significantly different for the $X$ and $Y$ polarizations, so fail
to cancel); and also by Galactic polarized straylight.
\item[Main Beam:] Cross-polarization is discussed in detail in the next 
section.
\item[Pointing:] Alignment errors between different horns corrupt
the reconstruction of the $(Q,U)$ vector at each pixel.
\item[Instrument Specific:] The principle effect is $1/f$ noise;
noise mismatch between $X$ and $Y$ radiometers may also cause minor
degradation in sensitivity. 
Some of these effects are discussed by Kaplan (these proceedings).
\item[Thermal, including internal straylight:]  
The responses of the instrument to  
many thermal effects are being modelled \citep{Mennella2001}.
Internal straylight
affects the polarization signal as for Galactic straylight, except that
only fluctuations are important.
\item[Calibration:]  Kaplan (these proceedings) discusses the
effect of inaccurate polarization calibration parameters.
\end{description}
In addition the scan strategy has unique effects on polarization, discussed
below.

Detailed studies of these effects are being made as part of the final 
design optimisation, to ensure that all contributions remain within the
top-level error budget.  However it is already clear that many will
not have a detectable impact on polarization.

As shown in detail in the following section, cross-polarization effects
from the main beam appear to be largely negligible, as are the different
responses to Stokes $I$ of the $X$ and $Y$ polarizations. 
Much larger differences are expected in the far sidelobes, 
but there the largest peaks turn out to be spillover past the subreflector
and the main mirror, and initial analysis suggests that the ${\cal I}_P$
response in these peaks differs by only about 10\%; this contributes
to $\delta g_X - \delta g_Y$ in terms of the analysis of the previous section.
An analysis of total-intensity Galactic straylight 
\citep{Burigana2001} showed an expected peak contamination of 
$\sim$4 $\mu$K
at high Galactic latitudes in the worst-case 30~GHz band, corresponding
to $\delta T < 1\, \mu$K in the $C_\ell$ spectrum. Therefore the polarized
signal should be less than $\sim$0.1 $\mu$K in the power spectrum, which
is just below the low-$\ell$ white noise at this frequency. 
At higher LFI frequencies, Galactic straylight
will be even lower, because of the falling spectrum of Galactic emission,
and also the increased forward gain.

Pointing errors produce error terms in the images proportional
to the local gradient. Typical gradients are of order the amplitude of 
structure on the scale of the beam, divided by the beamwidth. 
The pointing requirements for {\it Planck}
are set by HFI observations in total intensity, for which the ratio of
amplitude/beamwidth will be $>20$ times that for LFI observations of 
$Q$ or $U$.
so for our purposes the pointing is expected to be essentially perfect.

Instrument-specific and thermal effects mainly cause artefacts in the 
polarization
signal through the same mechanisms that affect the temperature signal. These
are among the hardest to deal with, and drive many aspects of the LFI and
{\em Planck} mission design. 
As noted earlier, suppression of these effects in the temperature signal
will automatically suppress them in the polarization signal as well,
but it is worth noting that often
(especially for thermal effects), they produce strongly correlated artefacts
in the two channels of each feed and so tend to cancel in the polarization 
signal.

\subsection{Cross Polarization}
\label{cross}

A radiometer operating at a single frequency receiving from a single direction
(i.e. a point source)
must couple perfectly to some purely polarized signal, 
that is, in terms of our Stokes response functions,
${\cal I}_P^2  = {\cal Q}_P^2 + {\cal U}_P^2 + {\cal V}_P^2$. In this
sense the concept of `cross-polarization', which implies that a system
always has some sensitivity to the `wrong' polarization, is rather
misleading.  In practice what is meant is that we'd like to build a
system sensitive only to $Q$ (say), but we find that ${\cal U}_P$ and
${\cal V}_P$ are non-zero.  Unwanted sensitivity to $V$ is straightforward:
it reduces the linear polarization efficiency $\epsilon_P$ but generally
creates no false positives, as $V \ll Q,U$. One can also lose efficiency
if $\phi$ varies with frequency or across the beam. However, non-zero
response to $U$ at the beam centre (or averaged over the beam) 
is best described as an error in $\phi$ rather than as cross polarization
in the sense usually intended.

We have been assessing the impact of these effects on the LFI, using
physical optics software written by V. Yurchenko \citep{Yur2001}
and comparing these
to simulations with the commercial GRASP8 package described by Villa
in his presentation. The two sets of results are in excellent agreement,
differing mainly through small differences in the assumed illumination
patterns of the feeds.  

{\it A priori} we do not expect the LFI beams to be very
well behaved in polarization as the feeds are located $3\dgr$ to $4\dgr$ away 
from the centre of the field of view, and the offset pseudo-Gregorian
design of the {\em Planck} telescope breaks circular symmetry.  However, 
the design maximises the effective field of view, while the feeds illuminate
the telescope with a strong edge taper, making near-in sidelobes extremely
low by the standards of normal radio telescopes; they are at $\lta -30$ dB
below the peak. Given the absence of sharp spikes in the CMB fluctuations, the 
total intensity beams are effectively Gaussian.

The most troublesome property of the beams is that they are
elliptical (axial ratio ranging from 1.14 to 1.39), 
with the horns in each `$Q,U$' matched pair differing in the orientation of the
ellipse by large angles. Thus deconvolution will be
needed to recover the $(Q,U)$ vector at each full-resolution sky
pixel; but for background measurements some degradation in resolution
is necessary to achieve adequate signal-to-noise in polarization, 
allowing resolution to be matched through linear techniques.

The total intensity  (${\cal I}_P$) beams for the $X$ and $Y$ channels
have effective areas identical to better than a few parts in $10^{5}$;
mismatches contribute to $\delta g$ in the analysis of the previous 
section, but acting only on spatial frequencies near the beamwidth.
On this scale amplitudes are $\sim 50$ $\mu$K, giving negligible artefacts
in the polarization signal. 

\begin{figure}
\resizebox{\columnwidth}{!}
{\includegraphics{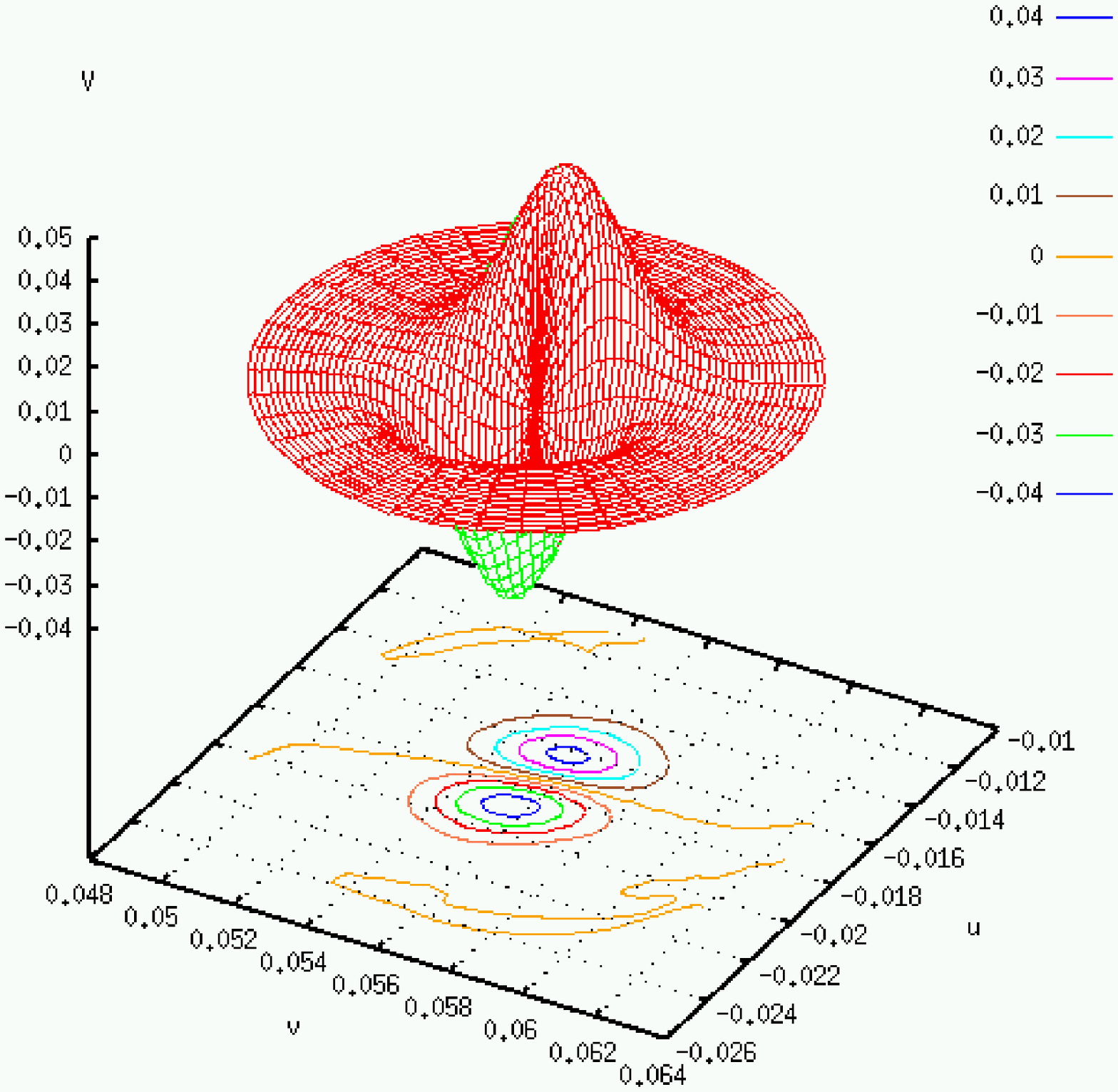}
 \includegraphics{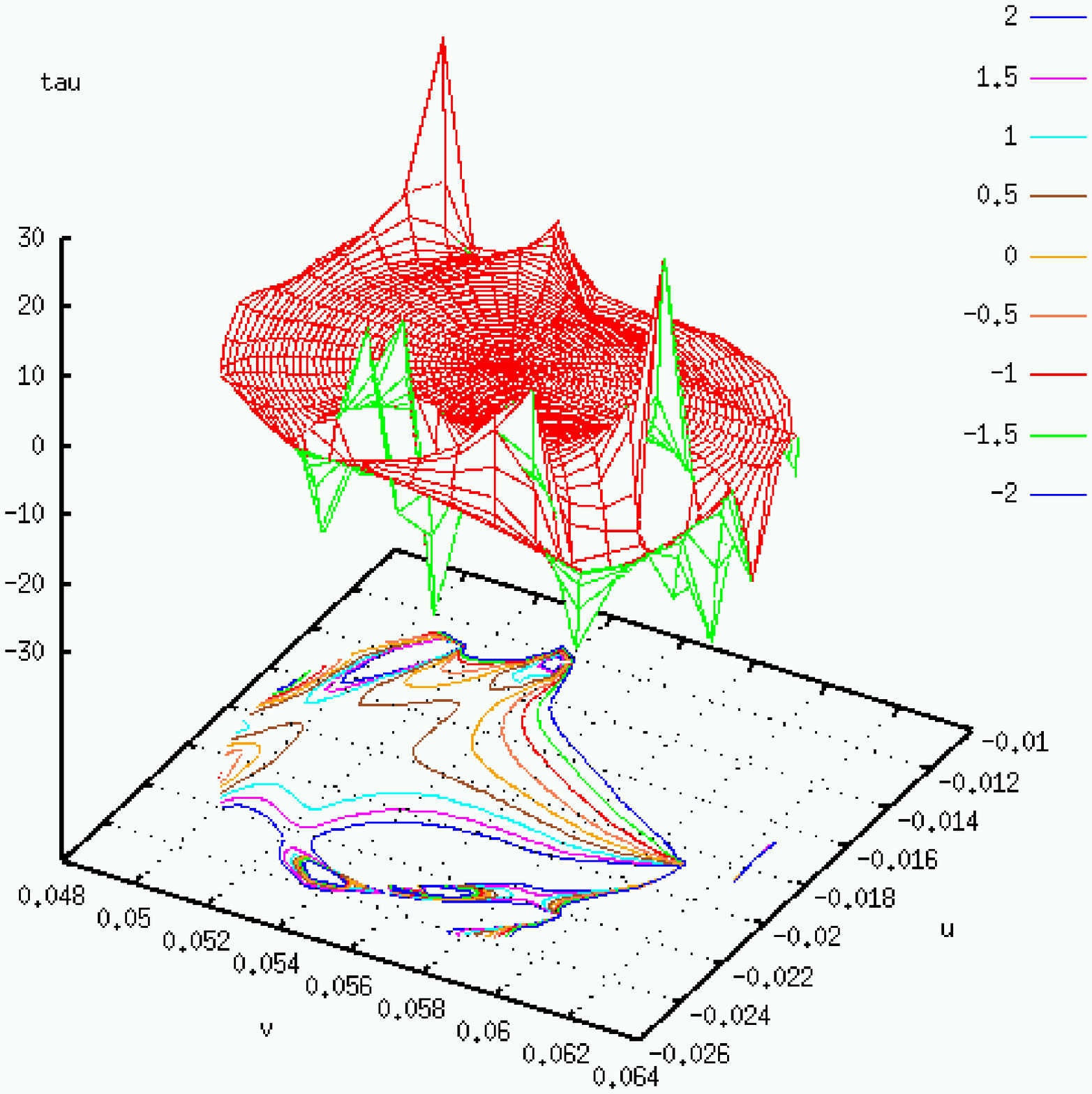}}
\caption{Simulations of the polarization response of a 100 GHz feed horn
(LFI-9). The coordinates ($u,v$) are direction cosines with respect to the
boresight direction. For comparison, the total intensity beam is centred at
($-$0.018, 0.056) and the $-$10 dB contour has a diameter of about 0.005 
in ($u,v$).
Left (a): Response to Stokes $V$, i.e. ${\cal V}_P$. Right (b): Variation
of orientation angle $\phi_P$, in degrees.}
\label{Xbeam}
\end{figure}

The major `unwanted' term turns out to be sensitivity to $V$, which
peaks at several percent of the main beam peak (Fig.~\ref{Xbeam}a)
The ${\cal V}_P$ beams always show a rather symmetric positive-negative
structure with the zero line passing almost through the peak of the
${\cal I}_P$ beam. Integrated over the beam,
 $|{\cal V}_P| / {\cal I}_P \lta 0.2$\% in all cases.
Averaged over the beam, the polarization efficiency ranges between 99.1\%
and 99.7\%.  If we choose coordinates so the wanted polarization
is $Q$, the ${\cal U_P}$ beam shows a similar structure to ${\cal V}_P$,
but with an even lower amplitude.
In this case the zero line goes through the beam centre by definition,
and the positive-negative structure is inevitable as there is a smooth
gradient of $\phi$ across the centre of the beam. Within the $-$10 dB contour
of the ${\cal I}_P$ beam, $\phi$ varies by up to about $2\dgr$. Large
deviations occur only near nulls in the sidelobe pattern (Fig.~\ref{Xbeam}b).
The effective $\phi$ evaluated from ${\cal Q_P}$ and ${\cal U_P}$ integrated
over the main beam differs from the value at the peak of 
${\cal I}_P$ by $< 0.15\dgr$.

We also find that the position angle in the far field rotates almost
perfectly with the E-field in the focal plane. As a result, if the
$X$ and $Y$ responses are orthogonal in the feed, the peak and effective
values of $\phi_X$ and $\phi_Y$ remain orthogonal in the far field to
within $\delta\phi \le 0.2\dgr$.

These results justify our claim that cross-polarization effects
are negligible for the LFI, at least at the pixel level.  Kaplan 
(these proceedings) has followed the effect of miscalibration of these
parameters through to the $C_\ell$ spectra. Here averaging down the thermal
noise makes residual polarization systematics more important, but comparison
of our results with his
findings suggest that even for $C_\ell$, the errors induced by the 
{\em Planck} telescope optics for the LFI horns will be almost undetectable;
presumably even more so for the more favourably placed HFI horns.

\begin{figure}
\resizebox{0.4\columnwidth}{!}
{\rotatebox{-90}{\includegraphics{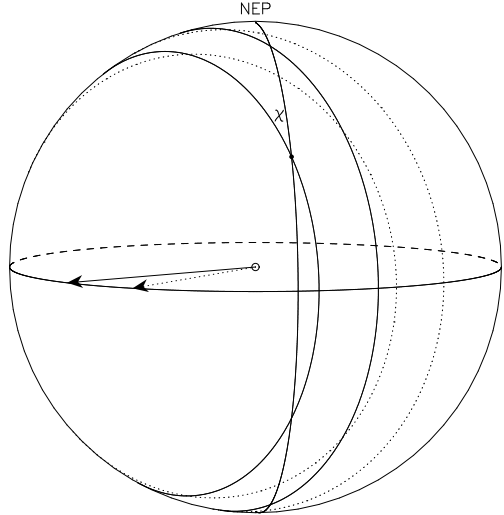}}
}
\caption{Geometry of scan paths. The arrows mark the direction of the
spacecraft spin axis at two different times. Scan circles from two horns
are shown (the difference in scan radius is exaggerated). 
NEP is the North Ecliptic Pole.
The angle $\chi$ is shown for one scan circle at the marked point. Note
that at
high latitudes scan circles through a given point intersect at significant
angles $\Delta\chi$.}
\label{sketch}
\end{figure}

\section{Feed Orientation and Scanning Strategy for the LFI}

If several measurements of components (not necessarily
orthogonal) of the ($Q, U$) vector are made, 
each with a Gaussian error distribution,
the 2D error distribution will in general be an elliptical Gaussian 
\citep{SA1999}.
Ideally we would like this distribution to be circular as this minimises
the area \citep{Couchot1999}; 
furthermore asymmetric errors will certainly complicate the
derivation of polarization $C_\ell$ spectra, 
and may lead to subtle biases, although no detailed
assessment has been made.

It is easy to see that if measurements are made with equally sensitive
detectors, with orientations $\phi$ such that $\{2\phi\}$ are 
evenly distributed
around a circle, then the error distribution will be circular. This is
satisfied by the normal set-up of two horns each with orthogonal feeds,
oriented at $45\dgr$ to each other; and also by the conventional arrangement
in optical polarimetry of $\phi = -60\dgr, 0\dgr, 60\dgr$.  Obviously,
combining several such sets with arbitrary offsets in $\phi$ will also yield
circular error distributions. 

The actual orientations measured on the sky is 
$\phi_{\rm sky} = \phi_S + \chi$, where
$\phi_S$ is the orientation of the radiometer polarization relative 
to the scan direction, and $\chi$ is the orientation of the scan circle
relative to the sky grid chosen to determine the zero of $\phi$. 
The default strategy for {\em Planck} is for the spin axis to point in the
anti-Sun direction towards the Ecliptic, which makes ecliptic meridians
a convenient reference direction (Fig.~\ref{sketch}).  The figure shows
that in this case 
$\chi$ is a function of ecliptic latitude, ranging from $0\dgr$
on the Ecliptic to $90\dgr$ at the maximum accessible latitude.

\begin{figure}
\resizebox{0.7\columnwidth}{!}{
    \rotatebox{-90}{
        \includegraphics{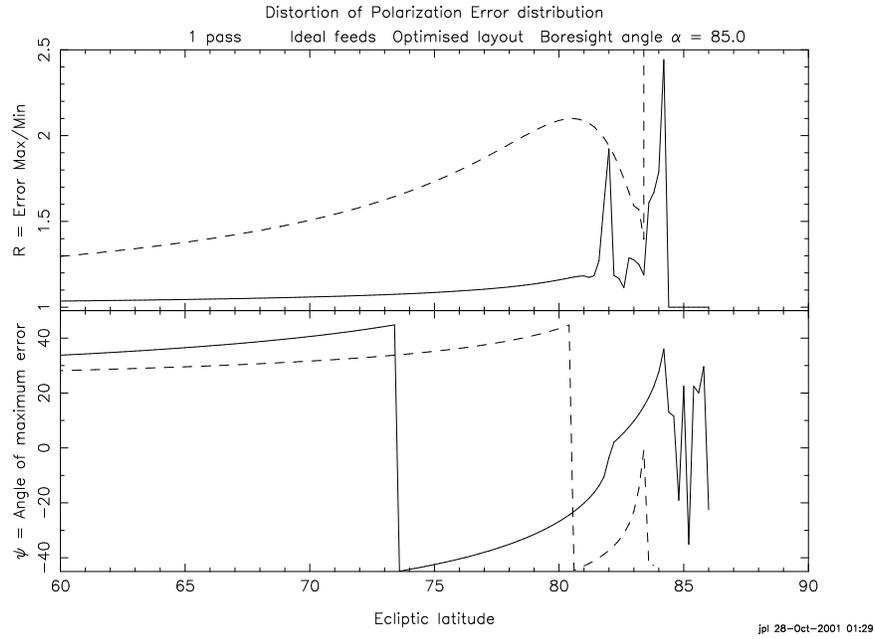}
    }
}
\caption{Axial ratio (top) and orientation (bottom) 
of the error ellipse as a function of ecliptic latitude, 
after one sky coverage, for the current LFI focal plane. Solid: 100~GHz:
Dashed: 44~GHz, assuming $\phi_S = 0\dgr, 60\dgr, -60\dgr$ for the three 
horns.}
\label{ellip}
\end{figure}

The radius of the LFI scan circles varies from $81\dgr$ to $89\dgr$.
Consequently the values of $\chi$ of different detectors measuring
the same sky pixel varies, especially near the ecliptic poles. The
LFI focal plane has been designed so that most matched pairs of feeds
are symmetrically placed on each side of the focal plane parallel
to the scan direction (barring small deviations of the spin axis
from nominal), so that they share the same scan circle and
hence the same relative orientation for all pixels. Unfortunately 
mechanical constraints prevent this arrangement for four of the sixteen
100 GHz feeds. These are arranged instead as a pair with 
$\phi_S = 90\dgr$ on one scan circle, and a pair with $\phi_S = 45\dgr$ on
another. At 44 GHz there are three feeds, two sharing a common
scan circle and the third on one several degrees larger. In these
cases some ellipticity in the $(Q,U)$ error distribution is inevitable.
Fig.~\ref{ellip} shows the effect on the error ellipse at high latitudes;
it is particularly large at 44~GHz because of the large
difference in radius between the two scan circles. A second coverage
improves the situation slightly since on the second pass pixels are
scanned by the other side of the scan circle, hence at angle $-\chi$.
Even so, we are left with highly elliptical error distributions
near the ecliptic poles, precisely the places where integration times
are longest and we have the best chance of detecting polarization at
high resolution.

\begin{figure}
\resizebox{0.8\columnwidth}{!}{
    \rotatebox{90}{
        \includegraphics*[5cm,1.2cm][18.5cm,26.8cm]{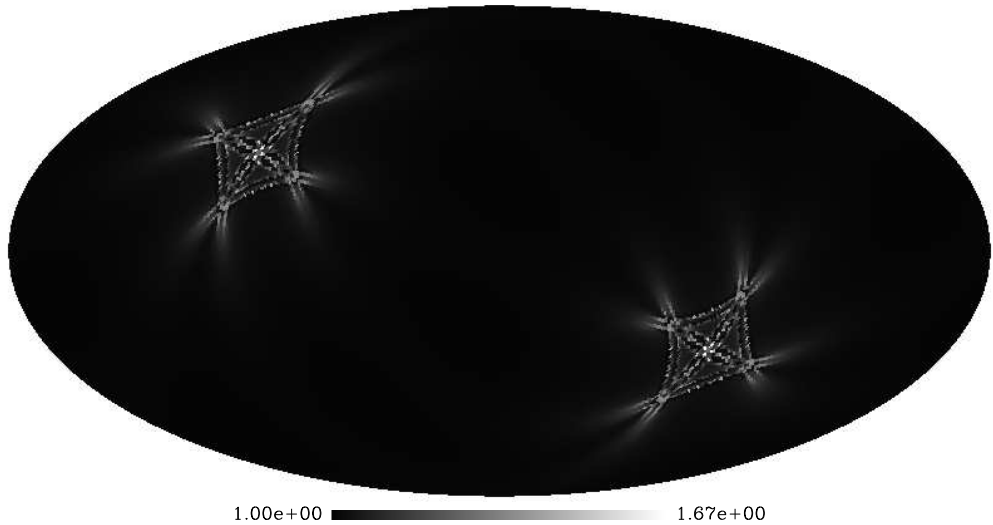}
    }
} 
\caption{Sky plot of error ellipticity at 100~GHz assuming a 
$5\dgr$ amplitude sinusoidal oscillation in latitude with
a wavelength of $90\dgr$ in longitude. The coordinate system is Galactic,
so that the structure around the ecliptic poles can clearly be seen.}
\label{hastie}
\end{figure}

To avoid this problem, and also holes in the coverage at the poles, the
spin axis must be moved off the Ecliptic. Several scan strategies are
under consideration, including a cycloidal path which maintains a constant
angle between the Sun and the spin axis, and a simple sinusoidal oscillation
in latitude. In these schemes the polarization error ellipse
depends on both latitude and longitude. Fig.~\ref{hastie}
shows an example at 100~GHz.  As far as polarization response
goes this option is a significant advance over the default strategy, with
rather circular error distributions up to the pole except at isolated
position on caustics,
and even here the axial ratio reaches only 1.7. Similar improvements are
seen at 44~GHz.

\section{Conclusions}
 
We have shown that in a differencing polarimeter the dominant systematics are
common to total intensity and polarization signals, except that some
cancellation can be expected in polarization.  As all LFI horns provide
a measurement of $I$ but each only provides one component of the $(Q,U)$
vector, the thermal noise in $Q$ and $U$ will be $\sqrt{2}$ larger than
in $I$. Thus we believe that if we can reach the goal of noise limited
sky maps in $I$, we should be able to do the same for $Q$ and $U$. 
Of course, $Q$ and $U$ will not be detected at full resolution in
most individual pixels, so the true test of our ability to do useful
polarization science is to follow the systematics through to the $C_\ell$
spectra, and much work is needed to complete this task.  But the low
level of polarization-specific systematics suggests that here again the
hard systematics to beat will be the same ones that affect total intensity,
and we know that these can be eliminated with high precision.

Finally we have shown that the polarization response, especially in the
most deeply surveyed regions near the ecliptic poles, can be substantially
improved by moving the spin axis away from the Ecliptic plane. Further
studies of alternative scan strategies will be made to help the project
decide on the best option.

\section{Acknowledgments}
We thank our colleagues in the {\em Planck} Systematics Effects Work Group
on polarization for useful discussions, especially Althea Wilkinson,
Fabrizio Villa, Jacques Delabrouille and Jean Kaplan. We acknowledge the
use of the HEALPix package (www.eso.org/science/healpix/).

\bibliographystyle{aipproc}
\bibliography{Planck}
\end{document}